\documentclass{ws-ijmpa}
\usepackage[compress]{cite}
\usepackage{graphicx,amsmath,float}
\usepackage{hyperref}

\let\svejk\v
\let\originalleft\left
\let\originalright\right
\renewcommand{\left}{\mathopen{}\mathclose\bgroup\originalleft}
\renewcommand{\right}{\aftergroup\egroup\originalright}

\DeclareSymbolFont{matha}{OML}{txmi}{m}{it}
\DeclareMathSymbol{\varv}{\mathord}{matha}{118}

\ifdefined\myframe
\renewenvironment{myframe}[2]{\section{#1}\begin{frame}{#2}\vspace{-10pt}}{\end{frame}} 
\else
\newenvironment{myframe}[2]{\section{#1}\begin{frame}{#2}\vspace{-10pt}}{\end{frame}}
\fi

\ifdefined\U
\renewcommand{\U}[1]{\underline{#1}}
\else
\newcommand{\U}[1]{\underline{#1}}
\fi

\ifdefined\UU
\renewcommand{\UU}[1]{\underline{\underline{#1}}}
\else
\newcommand{\UU}[1]{\underline{\underline{#1}}}
\fi

\ifdefined\mybullet
\renewcommand{\mybullet}{\vspace{2mm}\\$\bullet$ }
\else
\newcommand{\mybullet}{\vspace{2mm}\\$\bullet$ }
\fi

\ifdefined\bigXlohi
\renewcommand{\bigXlohi}[3]{\overset{#3}{\underset{#2}{\operatorname{\scalebox{#1}{$\mathsf X$}}}}}
\else
\newcommand{\bigXlohi}[3]{\overset{#3}{\underset{#2}{\operatorname{\scalebox{#1}{$\mathsf X$}}}}}
\fi

\ifdefined\bigXlo
\renewcommand{\bigXlo}[2]{\underset{#2}{\operatorname{\scalebox{#1}{$\mathsf X$}}}}
\else
\newcommand{\bigXlo}[2]{\underset{#2}{\operatorname{\scalebox{#1}{$\mathsf X$}}}}
\fi

\ifdefined\bigXhi
\renewcommand{\bigXhi}[2]{\overset{#2}{\operatorname{\scalebox{#1}{$\mathsf X$}}}}
\else
\newcommand{\bigXhi}[2]{\overset{#2}{\operatorname{\scalebox{#1}{$\mathsf X$}}}}
\fi

\ifdefined\bigX
\renewcommand{\bigX}[1]{\operatorname{\scalebox{#1}{$\mathsf X$}}}
\else
\newcommand{\bigX}[1]{\operatorname{\scalebox{#1}{$\mathsf X$}}}
\fi

\ifdefined\mybulletEQ
\renewcommand{\mybulletEQ}[1]{$\bullet$ {\bf #1}\vspace{1mm}\\}
\else
\newcommand{\mybulletEQ}[1]{$\bullet$ {\bf #1}\vspace{1mm}\\}
\fi

\ifdefined\fract
\renewcommand{\fract}[2]{{\textstyle \frac{#1}{#2}}}
\else
\newcommand{\fract}[2]{{\textstyle \frac{#1}{#2}}}
\fi

\ifdefined\rect
\renewcommand{\fract}[1]{{\textstyle \frac{1}{#1}}}
\else
\newcommand{\rect}[1]{{\textstyle \frac{1}{#1}}}
\fi

\ifdefined\fracd
\renewcommand{\fracd}[2]{{\displaystyle\frac{{#1}}{{#2}}}}
\else
\newcommand{\fracd}[2]{{\displaystyle\frac{{#1}}{{#2}}}}
\fi

\ifdefined\recd
\renewcommand{\recd}[1]{{\displaystyle\frac{1}{{#1}}}}
\else
\newcommand{\recd}[1]{{\displaystyle\frac{1}{{#1}}}}
\fi

\ifdefined\pdd
\renewcommand{\pdd}[2]{{\displaystyle\frac{\partial{#1}}{\partial{#2}}}}
\else
\newcommand{\pdd}[2]{{\displaystyle\frac{\partial{#1}}{\partial{#2}}}}
\fi

\ifdefined\pdt
\renewcommand{\pdt}[2]{{\textstyle\frac{\partial{#1}}{\partial{#2}}}}
\else
\newcommand{\pdt}[2]{{\textstyle\frac{\partial{#1}}{\partial{#2}}}}
\fi

\ifdefined\tdd
\renewcommand{\tdd}[2]{{\displaystyle\frac{d{#1}}{d{#2}}}}
\else
\newcommand{\tdd}[2]{{\displaystyle\frac{d{#1}}{d{#2}}}}
\fi

\ifdefined\tdt
\renewcommand{\tdt}[2]{{\textstyle{\frac{d{#1}}{d{#2}}}}}
\else
\newcommand{\tdt}[2]{{\textstyle{\frac{d{#1}}{d{#2}}}}}
\fi

\ifdefined\biggg
\renewcommand{\biggg}[1]{\scalebox{1.2}{\Bigg{#1}}}
\else
\newcommand{\biggg}[1]{\scalebox{1.2}{\Bigg{#1}}}
\fi

\ifdefined\Biggg
\renewcommand{\Biggg}[1]{\scalebox{1.4}{\Bigg{#1}}}
\else
\newcommand{\Biggg}[1]{\scalebox{1.4}{\Bigg{#1}}}
\fi

\ifdefined\Re

\renewcommand{\Re}{\operatorname{Re}}
\else
\newcommand{\Re}{\operatorname{Re}}
\fi

\ifdefined\Im

\renewcommand{\Im}{\operatorname{Im}}
\else
\newcommand{\Im}{\operatorname{Im}}
\fi

\ifdefined\Arch
\renewcommand{\Arch}{\operatorname{Ar\,ch}}
\else
\newcommand{\Arch}{\operatorname{Ar\,ch}}
\fi

\ifdefined\Arsh
\renewcommand{\Arsh}{\operatorname{Ar\,sh}}
\else
\newcommand{\Arsh}{\operatorname{Ar\,sh}}
\fi

\ifdefined\Arth
\renewcommand{\Arth}{\operatorname{Arth}}
\else
\newcommand{\Arth}{\operatorname{Arth}}
\fi

\ifdefined\ch
\renewcommand{\ch}{\operatorname{ch}}
\else
\newcommand{\ch}{\operatorname{ch}}
\fi

\ifdefined\sh
\renewcommand{\sh}{\operatorname{sh}}
\else
\newcommand{\sh}{\operatorname{sh}}
\fi

\ifdefined\th
\renewcommand{\th}{\operatorname{th}}
\else
\newcommand{\th}{\operatorname{th}}
\fi

\ifdefined\Ln
\renewcommand{\Ln}{\operatorname{Ln}}
\else
\newcommand{\Ln}{\operatorname{Ln}}
\fi

\ifdefined\tg
\renewcommand{\tg}{\operatorname{tg}}
\else
\newcommand{\tg}{\operatorname{tg}}
\fi

\ifdefined\ctg
\renewcommand{\ctg}{\operatorname{ctg}}
\else
\newcommand{\ctg}{\operatorname{ctg}}
\fi

\ifdefined\smallsum
\renewcommand{\smallsum}[2]{{\textstyle\sum\limits_{#1}^{#2}}}
\else
\newcommand{\smallsum}[2]{{\textstyle\sum\limits_{#1}^{#2}}}
\fi

\ifdefined\intl
\renewcommand{\intl}{\int\limits}
\else
\newcommand{\intl}{\int\limits}
\fi

\ifdefined\ointl
\renewcommand{\ointl}{\oint\limits}
\else
\newcommand{\ointl}{\oint\limits}
\fi

\ifdefined\soint
\renewcommand{\soint}[4]{\hspace{#1}\oint_{#3}^{#4}\hspace{#2}}
\else
\newcommand{\soint}[4]{\hspace{#1}\oint_{#3}^{#4}\hspace{#2}}
\fi

\ifdefined\sointl
\renewcommand{\sointl}[4]{\hspace{#1}\oint\limits_{#3}^{#4}\hspace{#2}}
\else
\newcommand{\sointl}[4]{\hspace{#1}\oint\limits_{#3}^{#4}\hspace{#2}}
\fi

\ifdefined\shoint
\renewcommand{\shoint}[3]{\hspace{#1}\oint^{#3}\hspace{#2}}
\else
\newcommand{\shoint}[3]{\hspace{#1}\oint^{#3}\hspace{#2}}
\fi

\ifdefined\shointl
\renewcommand{\shointl}[3]{\hspace{#1}\oint\limits^{#3}\hspace{#2}}
\else
\newcommand{\shointl}[3]{\hspace{#1}\oint\limits^{#3}\hspace{#2}}
\fi

\ifdefined\sloint
\renewcommand{\sloint}[3]{\hspace{#1}\oint_{#3}\hspace{#2}}
\else
\newcommand{\sloint}[3]{\hspace{#1}\oint_{#3}\hspace{#2}}
\fi

\ifdefined\slointl
\renewcommand{\slointl}[3]{\hspace{#1}\oint\limits_{#3}\hspace{#2}}
\else
\newcommand{\slointl}[3]{\hspace{#1}\oint\limits_{#3}\hspace{#2}}
\fi

\ifdefined\sint
\renewcommand{\sint}[4]{\hspace{#1}\int_{#3}^{#4}\hspace{#2}}
\else
\newcommand{\sint}[4]{\hspace{#1}\int_{#3}^{#4}\hspace{#2}}
\fi

\ifdefined\sintl
\renewcommand{\sintl}[4]{\hspace{#1}\int\limits_{#3}^{#4}\hspace{#2}}
\else
\newcommand{\sintl}[4]{\hspace{#1}\int\limits_{#3}^{#4}\hspace{#2}}
\fi

\ifdefined\slint
\renewcommand{\slint}[3]{\hspace{#1}\int_{#3}\hspace{#2}}
\else
\newcommand{\slint}[3]{\hspace{#1}\int_{#3}\hspace{#2}}
\fi

\ifdefined\slintl
\renewcommand{\slintl}[3]{\hspace{#1}\int\limits_{#3}\hspace{#2}}
\else
\newcommand{\slintl}[3]{\hspace{#1}\int\limits_{#3}\hspace{#2}}
\fi

\ifdefined\integrated
\renewcommand{\integrated}[3]{\left\{{#1}\right\}\left.\vphantom{#1}\right|_{#2}^{#3}}
\else
\newcommand{\integrated}[3]{\left\{{#1}\right\}\left.\vphantom{#1}\right|_{#2}^{#3}}
\fi

\ifdefined\pd
\renewcommand{\pd}[2]{\frac{\partial{#1}}{\partial{#2}}}
\else
\newcommand{\pd}[2]{\frac{\partial{#1}}{\partial{#2}}}
\fi

\ifdefined\rec
\renewcommand{\rec}[1]{\frac{1}{#1}}
\else
\newcommand{\rec}[1]{\frac{1}{#1}}
\fi

\ifdefined\gvec
\renewcommand{\gvec}[1]{\mbox{\boldmath${#1}$}}
\else
\newcommand{\gvec}[1]{\mbox{\boldmath${#1}$}}
\fi

\ifdefined\cvec
\renewcommand{\cvec}[1]{\mbox{\boldmath${#1}$}}
\else
\newcommand{\cvec}[1]{\mbox{\boldmath${#1}$}}
\fi

\ifdefined\td
\renewcommand{\td}[2]{\frac{d{#1}}{d{#2}}}
\else
\newcommand{\td}[2]{\frac{d{#1}}{d{#2}}}
\fi

\ifdefined\md
\renewcommand{\md}[2]{\frac{\mathrm{d}{#1}}{\mathrm{d}{#2}}}
\else
\newcommand{\md}[2]{\frac{\mathrm{d}{#1}}{\mathrm{d}{#2}}}
\fi

\ifdefined\z
\renewcommand{\z}[1]{\left({#1}\right)}
\else
\newcommand{\z}[1]{\left({#1}\right)}
\fi

\ifdefined\ae
\renewcommand{\ae}[1]{\left|{#1}\right|}
\else
\newcommand{\ae}[1]{\left|{#1}\right|}
\fi

\ifdefined\sz
\renewcommand{\sz}[1]{\left[{#1}\right]}
\else
\newcommand{\sz}[1]{\left[{#1}\right]}
\fi

\ifdefined\kz
\renewcommand{\kz}[1]{\left\{{#1}\right\}}
\else
\newcommand{\kz}[1]{\left\{{#1}\right\}}
\fi

\ifdefined\B
\renewcommand{\B}[1]{\mathbb{#1}}
\else
\newcommand{\B}[1]{\mathbb{#1}}
\fi

\ifdefined\m
\renewcommand{\m}[1]{\mathrm{#1}}
\else
\newcommand{\m}[1]{\mathrm{#1}}
\fi

\ifdefined\tn
\renewcommand{\tn}[1]{\textnormal{#1}}
\else
\newcommand{\tn}[1]{\textnormal{#1}}
\fi

\ifdefined\o
\renewcommand{\o}[1]{\operatorname{#1}}
\else
\newcommand{\o}[1]{\operatorname{#1}}
\fi

\ifdefined\c
\renewcommand{\c}[1]{\mathcal{#1}}
\else
\newcommand{\c}[1]{\mathcal{#1}}
\fi

\ifdefined\f
\renewcommand{\f}[1]{\mathfrak{#1}}
\else
\newcommand{\f}[1]{\mathfrak{#1}}
\fi

\ifdefined\bs
\renewcommand{\bs}[1]{\boldsymbol{#1}}
\else
\newcommand{\bs}[1]{\boldsymbol{#1}}
\fi

\ifdefined\v
\renewcommand{\v}[1]{\mathbf{#1}}
\else
\newcommand{\v}[1]{\mathbf{#1}}
\fi

\ifdefined\Eq
\renewcommand{\Eq}[1]{Eq.~(\ref{#1})}
\else
\newcommand{\Eq}[1]{Eq.~(\ref{#1})}
\fi

\ifdefined\Eqs
\renewcommand{\Eqs}[2]{Eqs.~(\ref{#1}) and (\ref{#2})}
\else
\newcommand{\Eqs}[2]{Eqs.~(\ref{#1}) and (\ref{#2})}
\fi

\ifdefined\a
\renewcommand{\a}[1]{\aref({#1})}
\else
\newcommand{\a}[1]{\aref({#1})}
\fi

\ifdefined\A
\renewcommand{\A}[1]{\Aref({#1})}
\else
\newcommand{\A}[1]{\Aref({#1})}
\fi

\ifdefined\r

\renewcommand{\r}[1]{(\ref{#1})}
\else
\newcommand{\r}[1]{(\ref{#1})}
\fi

\ifdefined\comm
\renewcommand{\comm}[2]{\left[{#1},{#2}\right]}
\else
\newcommand{\comm}[2]{\left[{#1},{#2}\right]}
\fi

\ifdefined\follows
\renewcommand{\follows}{\quad\Rightarrow\quad}
\else
\newcommand{\follows}{\quad\Rightarrow\quad}
\fi

\ifdefined\Follows
\renewcommand{\Follows}{\qquad\Rightarrow\qquad}
\else
\newcommand{\Follows}{\qquad\Rightarrow\qquad}
\fi

\ifdefined\followse
\renewcommand{\followse}{\quad\Rightarrow}
\else
\newcommand{\followse}{\quad\Rightarrow}
\fi

\ifdefined\bfollows
\renewcommand{\bfollows}{\Rightarrow\quad}
\else
\newcommand{\bfollows}{\Rightarrow\quad}
\fi

\ifdefined\equivalent
\renewcommand{\equivalent}{\quad\Leftrightarrow\quad}
\else
\newcommand{\equivalent}{\quad\Leftrightarrow\quad}
\fi

\ifdefined\Equivalent
\renewcommand{\Equivalent}{\qquad\Leftrightarrow\qquad}
\else
\newcommand{\Equivalent}{\qquad\Leftrightarrow\qquad}
\fi

\ifdefined\obs
\renewcommand{\obs}[1]{\left\langle{#1}\right\rangle}
\else
\newcommand{\obs}[1]{\left\langle{#1}\right\rangle}
\fi

\ifdefined\ket
\renewcommand{\ket}[1]{\left|{#1}\right\rangle}
\else
\newcommand{\ket}[1]{\left|{#1}\right\rangle}
\fi

\ifdefined\bra
\renewcommand{\bra}[1]{\left\langle{#1}\right|}
\else
\newcommand{\bra}[1]{\left\langle{#1}\right|}
\fi

\ifdefined\braket
\renewcommand{\braket}[2]{\left<#1\vphantom{#2}\right|\left.#2\vphantom{#1}\right>}
\else
\newcommand{\braket}[2]{\left<#1\vphantom{#2}\right|\left.#2\vphantom{#1}\right>}
\fi

\ifdefined\ketbra
\renewcommand{\ketbra}[2]{\left|#1\vphantom{#2}\right>\left<#2\vphantom{#1}\right|}
\else
\newcommand{\ketbra}[2]{\left|#1\vphantom{#2}\right>\left<#2\vphantom{#1}\right|}
\fi

\ifdefined\scalprod
\renewcommand{\scalprod}[2]{\left(#1\vphantom{#2}\right|\left.#2\vphantom{#1}\right)}
\else
\newcommand{\scalprod}[2]{\left(#1\vphantom{#2}\right|\left.#2\vphantom{#1}\right)}
\fi

\ifdefined\fixmatrix
\renewcommand{\fixmatrix}[2]{\left(\begin{array}{*{9}{@{}>{\centering\arraybackslash $}m{#1}<{$ }@{}}}#2\end{array}\right)}
\else
\newcommand{\fixmatrix}[2]{\left(\begin{array}{*{9}{@{}>{\centering\arraybackslash $}m{#1}<{$ }@{}}}#2\end{array}\right)}
\fi

\ifdefined\fixgausselim
\renewcommand{\fixgausselim}[4]{\left(\hspace{-1mm}\begin{array}{*{9}{@{}>{\centering\arraybackslash $}m{#1}<{$ }@{}}}#3\end{array}\vphantom{\begin{array}{*{100}c}#4\end{array}}\hspace{-1mm}\right|\hspace{-1mm}\left.\begin{array}{*{9}{@{}>{\centering\arraybackslash $}m{#2}<{$ }@{}}}#4\end{array}\vphantom{\begin{array}{*{100}c}#3\end{array}}\right)}
\else
\newcommand{\fixgausselim}[4]{\left(\hspace{-1mm}\begin{array}{*{9}{@{}>{\centering\arraybackslash $}m{#1}<{$ }@{}}}#3\end{array}\vphantom{\begin{array}{*{100}c}#4\end{array}}\hspace{-1mm}\right|\hspace{-1mm}\left.\begin{array}{*{9}{@{}>{\centering\arraybackslash $}m{#2}<{$ }@{}}}#4\end{array}\vphantom{\begin{array}{*{100}c}#3\end{array}}\right)}
\fi

\ifdefined\gausselim
\renewcommand{\gausselim}[2]{\left(\begin{matrix}#1\end{matrix}\vphantom{\begin{matrix}#2\end{matrix}}\hspace{1mm}\right|\left.\begin{matrix}#2\end{matrix}\vphantom{\begin{matrix}#1\end{matrix}}\right)}
\else
\newcommand{\gausselim}[2]{\left(\begin{matrix}#1\end{matrix}\vphantom{\begin{matrix}#2\end{matrix}}\hspace{1mm}\right|\left.\begin{matrix}#2\end{matrix}\vphantom{\begin{matrix}#1\end{matrix}}\right)}
\fi

\ifdefined\distreff
\renewcommand{\distreff}[2]{\big({#1}\big|{#2}\big)}
\else
\newcommand{\distreff}[2]{\big({#1}\big|{#2}\big)}
\fi

\ifdefined\matrixel
\renewcommand{\matrixel}[3]{\left<#1\vphantom{#2#3}\right|#2\left|#3\vphantom{#1#2}\right>} 
\else
\newcommand{\matrixel}[3]{\left<#1\vphantom{#2#3}\right|#2\left|#3\vphantom{#1#2}\right>} 
\fi

\ifdefined\uplow 
\renewcommand{\uplow}[3]{{{#1}^{#2}_{}\hspace{-1.5pt}}_{#3}}
\else
\newcommand{\uplow}[3]{{{#1}^{#2}_{}\hspace{-1.5pt}}_{#3}}
\fi

\ifdefined\lowup
\renewcommand{\lowup}[3]{{{#1}_{#2}^{}\hspace{-1.5pt}}^{#3}}
\else
\newcommand{\lowup}[3]{{{#1}_{#2}^{}\hspace{-1.5pt}}^{#3}}
\fi

\ifdefined\am
\renewcommand{\am}{{\hat{a}^{\vphantom\dagger}}}
\else
\newcommand{\am}{{\hat{a}^{\vphantom\dagger}}}
\fi

\ifdefined\ap
\renewcommand{\ap}{{\hat{a}^\dagger}}
\else
\newcommand{\ap}{{\hat{a}^\dagger}}
\fi

\ifdefined\bm
\renewcommand{\bm}{{\hat{b}^{\vphantom\dagger}}}
\else
\newcommand{\bm}{{\hat{b}^{\vphantom\dagger}}}
\fi

\ifdefined\bp
\renewcommand{\bp}{{\hat{b}^\dagger}}
\else
\newcommand{\bp}{{\hat{b}^\dagger}}
\fi

\ifdefined\arctg
\renewcommand{\arctg}{\operatorname{arctg}}
\else
\newcommand{\arctg}{\operatorname{arctg}}
\fi

\begin{document}
\markboth{E. Árpási, M. I. Nagy}{Investigation of the pion pair-source in heavy-ion collisions}

%
\catchline{}{}{}{}{}
%

\title{Multi-dimensional investigation of the pion pair-source in heavy-ion collisions with EPOS}
\author{Emese Árpási}
\address{Institute of Physics, ELTE E\"otv\"os Lor\'and University, P\'azm\'any P\'eter s\'et\'any 1/A\\ Budapest, H-1117, Hungary\\ earpasi@student.elte.hu}
\author{M\'arton I. Nagy}
\address{Institute of Physics, ELTE E\"otv\"os Lor\'and University, P\'azm\'any P\'eter s\'et\'any 1/A\\ Budapest, H-1117, Hungary\\ nmarci@elte.hu}
\maketitle


\begin{abstract}

The purpose of femtoscopy is to unveil the space-time characteristics of heavy-ion collisions as mirrored in the correlation function of identical particles. We aim to augment such research by investigating the source function of particle creation based on a modern event generator, the EPOS package. It simulates high energy nuclear collisions and is proven to reliably reproduce the characteristics of real collisions. The source function has been almost always assumed to be Gaussian, but lately it is experimentally found that L\'evy-stable distributions provide a much more acceptable description. This paper focuses on the source shape that is reconstructed in a femtoscopic measurement, in particular, the event-by-event geometry as well as the deviation from spherical symmetry of the source in the EPOS simulation. Such an investigation tests some scenarios of the appearance of non-Gaussian source functions. We find that in such a realistic simulation, non-Gaussian sources indeed appear, even in a three-dimensional setting and without event averaging. However, our reconstructed L\'evy exponent (responsible for the long-range behavior of the source) is systematically larger than that observed in experiment; a discrepancy that might hint at further significance of this exponent.
\keywords{heavy-ion collisions, femtoscopy, non-Gaussian source functions}
\end{abstract}
\ccode{PACS numbers: 25.75}

\section{Introduction}

One of the main goals of heavy-ion physics is to study the strongly coupled Quark-Gluon-Plasma (sQGP) created in high energy heavy-ion collisions. The discovery of this deconfined state of matter at the RHIC accelerator~\cite{PHENIX:2004vcz,STAR:2005gfr,PHOBOS:2004zne,BRAHMS:2004adc} was accomplished by an assortment of measurements techniques and investigation methods. One of them is femtoscopy, the reconstruction of the probability distribution (source function) of particle production from the correlation function of identical particles. Indeed, this is almost the only way to study space-time geometries on such femtometer scale.

In a femtoscopic investigation, one usually employs a model function for the source function, and reconstructs its parameters from fits of the theoretically calculated correlation functions to the measured ones. The usual such model function was assumed to be Gaussian for a long time (see e.g. Ref.~\cite{PHENIX:2004yan}), but later experimental results suggested a power-law decay at large distances~\cite{PHENIX:2006nml,PHENIX:2007grx}. Recently, the assumption of L\'evy shapes (indeed encompassing power-law-like behavior at large distances) have become successful in experimental reconstruction of the source shape~\cite{PHENIX:2017ino,Kincses:2024sin,CMS:2023xyd,NA61SHINE:2023qzr,PHENIX:2024vjp,Csanad:2024hva}. However, much of the physical implications of this (even the fact that Gaussian shapes are inadequate to describe the source) remain to be understood. In particular, a concern was raised~\cite{Cimerman:2019hva} that maybe event averaging alone (inherent in experimental femtoscopic anaylses) could be responsible for the observed non-Gaussianity. Still assuming spherical symmetry, it was shown~\cite{Kincses:2022eqq,Korodi:2022ohn} that L\'evy functions as source shapes appear even in individual events. The purpose of the present work is to generalize this investigation to the case when spherical symmetry is not assumed, to see what role directional averaging plays in the L\'evy-like behavior of the source.

The structure of the paper is as follows. In Section~\ref{s:levy} we introduce femtoscopic correlation functions, L\'evy distributions, and detail various methods of practical calculations of such functions. In Section~\ref{s:application} we show an example for the application of these distributions in realistic simulations of heavy-ion collisions, the EPOS model. It turns out that L\'evy-stable distributions are needed for an acceptable description of the observed sources, even without the assumption of spherical symmetry, and on an event-by-event basis. Finally, Section~\ref{s:summary} summarizes our results.

\section{L\'evy-type source functions in femtoscopy}\label{s:levy}

\subsection{Femtoscopic correlation functions}\label{ss:femtoscopy}

Femtoscopy in high energy heavy-ion physics is to study the processes happening in the femtometer scale. It relies on the fact that correlated production of identical particles is governed by their quantum mechanical indistinguishability (as first explained by G. Goldhaber et al.~\cite{Goldhaber:1960sf}; for a recent review, see Refs.~\cite{Csorgo:1999sj,Lisa:2005dd}). Such correlations are thus often called Bose-Einstein correlations (in case of bosons, e.g. charged pions). Their basic expression is the Yano-Koonin formula~\cite{Yano:1978gk}:
\begin{align}
C(\v q)=\frac{\int d^4x D(x,\v K)|\Psi_{k_1,k_2}(x_1,x_2)|^2}{\int d^4x D(x,\v K)},
\end{align}
where $p_1$, $p_2$ are the particles momenta, $q\,{\equiv}\,p_1{-}p_2\,{\equiv}\,(q^0,\v q)$ is the relative four-momentum, $x\,{\equiv}\,x_1{-}x_2$ is the relative four-coordinate, $\v K\,{\equiv}\,\rec2(\v p_1{+}\v p_2)$ is the pair average momentum, $\Psi_{k_1,k_2}(x_1,x_2)$ is the pairwise wave-function of the particles and $D(x,\textbf{K})$ is the pair source distribution. This latter quantity is the auto-convolution of the particle production source function $S(X,\v p)$, and is thus expressed as
\begin{align}
D(x,\v K)=\sint{-2pt}{-3pt}{}{}d^4X\,S\big(X{+}\fract x2,\v K\big)S\big(X{-}\fract x2,\v K\big).
\label{Dcc}
\end{align}
Because $|\Psi_{k_1,k_2}(x_1,x_2)|^2$ contains interference terms (owing to the symmetrization in case of bosons), one gets a non-trivial correlation from the above formulae; for example, neglecting final state interactions, $C(\v q)\,{-}\,1$ becomes the Fourier transform of the pairwise source $D(x)$. It is thus this $D(x)$ that can be measured experimentally from correlations, and so it is $D(x)$ that we aim to investigate using event generators. As stated above, this was assumed to be a Gaussian for a long time, and while this turned out to be unacceptable against precision data, a fairly good description was achieved using L\'evy distributions. These were introduced in high energy physics femtoscopy in Ref.~\cite{Csorgo:2003uv}. Some possible physical scenarios that lead to such distributions are critical phenomena, anomalous diffusion, or jet fragmentation~\cite{Csorgo:2005it,Csanad:2007fr,Csorgo:2004sr}. It is thus worthwhile to investigate the appearance of L\'evy distributions in simulations. 

\subsection{L\'evy functions: definitions}

The Lévy function is a continuous probability distribution. A general form of the function in three dimensions admits four parameters; its expression is
\begin{align}
\c L^{(3D)}(\v r,R_x,R_y,R_z,\alpha) = \sint{-2pt}{-3pt}{}{}\frac{d^3q}{(2\pi)^3}\,e^{i\v q\v r}\m{exp}\big({-}\rect2|q_x^2R_x^2{+}q_y^2R_y^2{+}q_z^2R_z^2|^{\alpha/2}\big),
\label{eq:levy3d}
\end{align}
where the $R_x$, $R_y$ and $R_z$ parameters, called L\'evy scale parameters, describe the geometric sizes in the principal directions, while the $\alpha$ parameter is the
so-called L\'evy index. It is constrained into the $0\,{<}\,\alpha\,{\leq}\,2$ domain. We also introduce the 1 dimensional projection (in the notation already exploiting that it does not depend on the scale parameters in the directions that are integrated out) as 
\begin{align}
\c L^{(1D)}(r_x,R_x,\alpha)=\sint{-2pt}{-12pt}{-\infty}\infty dr_y\sint{-2pt}{-12pt}{-\infty}\infty dr_z\,\c L^{(3D)}(r_x,r_y,r_z,R_x,R_y,R_z,\alpha),
\end{align}
whose expression as a Fourier transform is then
\begin{align}
\c L^{(1D)}(r_x,R_x,\alpha)=\rec{2\pi}\sint{-2pt}{-12pt}{-\infty}\infty dq_x\,e^{ir_xq_x}\m{exp}\big({-}\rect2|q_xR_x|^\alpha\big).
\label{eq:levy1d}
\end{align}
These functions are normalized to 1:
\begin{align}
\sint{-2pt}{-3pt}{}{}d^3\v r\,\c L^{(3D)}(\v r,R_x,R_y,R_z,\alpha) = \sint{-2pt}{-12pt}{-\infty}\infty dr_x\,\c L^{(1D)}(r_x,R_x,\alpha) = 1.
\end{align}
By rescaling the integration variable in \r{eq:levy3d} and \r{eq:levy1d} it is easy to see that
\begin{align}
&\c L^{(3D)}(r_x,r_y,r_z,R_x,R_y,R_z,\alpha)=\rec{R_xR_yR_z}\c L^{(3D)}\bigg(\frac{r_x}{R_x},\frac{r_y}{R_y},\frac{r_z}{R_z},1,1,1,\alpha\bigg),\\
&\c L^{(1D)}(r_x,R_x,\alpha)=\rec{R_x}\c L^{(1D)}\bigg(\frac{r_x}{R_x},1,\alpha\bigg),
\end{align}
so it is enough to do the calculations with $R_x\,{=}\,1$ and $R_k\,{=}\,1$ ($k\,{=}\,x,y,z$), respectively. With $R_k\,{=}\,1$ even the 3D form is spherically symmetric, and the solid angle integral can be performed easily. It is thus convenient to use the re-scaled functions
\begin{align}
\label{e:l13ddef}
\c L_{3D}(r,\alpha) = \rec{2\pi^2}\sint{-2pt}{-12pt}0\infty dq\,q^2\frac{\sin(qr)}{qr}e^{-\rect2q^\alpha},\qquad
\c L_{1D}(r,\alpha) = \rec\pi\sint{-2pt}{-12pt}0\infty dq\,\cos(qr)e^{-\rect2q^\alpha}.
\end{align}
With these, owing to spherical symmetry, we thus have
\begin{align}
&\c L^{(3D)}(r_x,r_y,r_z,R_x,R_y,R_z,\alpha)=\rec{R_xR_yR_z}\c L_{3D}\bigg(\sqrt{\fract{r_x^2}{R_x^2}{+}\fract{r_y^2}{R_y^2}{+}\fract{r_z^2}{R_z^2}},\alpha\bigg),\\
&\c L^{(1D)}(r_x,R_x,\alpha)=\rec{R_x}\c L_{1D}\Big(\frac{r_x}{R_x},\alpha\Big).
\end{align}
We note that the values at $r{=}0$ can be expressed with the gamma function $\Gamma(x)$ as
\begin{align}
\c L_{3D}(r{=}0,\alpha) = \frac{2^{3/\alpha}}{2\pi^2\alpha}\Gamma\Big(\frac3\alpha\Big),\qquad
\c L_{1D}(r{=}0,\alpha) = \frac{2^{1/\alpha}}{\pi\alpha}\Gamma\Big(\rec\alpha\Big).
\end{align}
In the case of $\alpha{=}2$, the L\'evy distribution becomes a Gaussian:
\begin{align}
\c L_{3D}(r,\alpha{=}2) = (2\pi)^{-3/2}e^{-r^2/2}, \quad\qquad \c L_{1D}(r,\alpha{=}2) = (2\pi)^{-1/2}e^{-r^2/2};
\end{align}
while for $0\,{<}\,\alpha\,{<}\,2$, the function exhibits a power-law decay at large $r$ (see below). There is a particular case of interest, $\alpha{=}1$, when it becomes a Cauchy distribution: 
\begin{align}
\c L_{3D}(r,\alpha{=}1) = \frac8{\pi^2}\rec{(1{+}4r^2)^2}, \qquad \c L_{1D}(r,\alpha{=}1) = \frac2\pi\rec{1{+}4r^2}.
\end{align}
In other cases, there are no such simple analytical forms; some practicalities of this are detailed below in Sec.~\ref{ss:levy:calc}. We note that if $R_x=R_y=R_z$, then the L\'evy distribution retains the stability property against convolutions, known to be true for Gaussians. This is the reason why L\'evy distributions are expected to appear in many scenarios. In particular, if the source function is such a L\'evy distribution, then so will be the pair distribution, with a modified radius:
\begin{equation}
S(\v r)=\c L(\v r,R,\alpha) \Follows D(\v r)=\c L(\v r,2^{\rec\alpha}R,\alpha),
\end{equation}
where the shorthand $R$ was used here for $R_x\,{=}\,R_y\,{=}\,R_z$, and the pair transverse momentum $\v K$ (on which $R$ and $\alpha$ might depend) was suppressed in the notation.

\subsection{L\'evy functions: calculational methods}\label{ss:levy:calc}

Our study requires the fast and effective calculation of the L\'evy distribution functions $\c L_{1D}$ and $\c L_{3D}$. Much work has been done in this area (see e.g.~Refs.~\cite{Levy,Levy:old,Levy:num}); this subsection contains a review of some of such methods (possibly with some improvements on older results; without claim of originality nevertheless). We concentrate on the case relevant four our studies, $1\,{\le}\,\alpha\,{\le}\,2$.

A simple series expansion of the integrands of Eq.~\r{e:l13ddef} in terms of $r$ yields
\begin{align}
&\c L_{3D}(r,\alpha) = \frac{2^{3/\alpha}}{2\pi^2\alpha}\sum_{p=0}^\infty\big(2^{\rec\alpha}r\big)^{2p}\frac{(-1)^p}{(2p{+}1)!}\Gamma\Big(\frac{2p{+}3}\alpha\Big),\\
&\c L_{1D}(r,\alpha) = \frac{2^{1/\alpha}}{\pi\alpha}\sum_{p=0}^\infty\big(2^{\rec\alpha}r\big)^{2p}\frac{(-1)^p}{(2p)!}\Gamma\Big(\frac{2p{+}1}\alpha\Big).
\end{align}
These series are convergent for all $r\,{\ge}\,0$; however, practical use is limited by numerical instability of the oscillating terms at larger $r$ values.

On the other hand, we can rewrite the defining \r{e:l13ddef} integrals as
\begin{align}
\c L_{3D}(r,\alpha) = \rec{2\pi^2r}\m{Im}\sint{-2pt}{-12pt}0\infty dq\,qe^{iqr-\rect2q^\alpha}, \qquad 
\c L_{1D}(r,\alpha) = \rec\pi\m{Re}\sint{-2pt}{-12pt}0\infty dq\,e^{iqr-\rect2q^\alpha}.
\label{e:l13deiqr}
\end{align}
Substituting $q\,{\to}\,t$, $q\,{=}\,\rec ri^{1/\alpha}t$ (i.e. turning the integration path on the complex $q$ plane by an angle of $\frac\pi{2\alpha}$), and using for the $e^{it^\alpha/2r^\alpha}$ factor so obtained the identity valid for $x\,{\in}\,\B R^+$ and for any $N{\in}\B N_0^+$, $\big|e^{ix}\,{-}\,\sum_{n=0}^N\rec{n!}(ix)^n\big|\,{\le}\,\rec{(N{+}1)!}x^{N+1}$, we get
\begin{align}
\label{e:l3dasymp}
&\c L_{3D}(r,\alpha) = \rec{2\pi^2r^3}\bigg[\sum_{n=1}^N(-1)^{n+1}\m{sin}\Big(\frac{n\pi\alpha}2\Big)\frac{\Gamma(\alpha n{+}2)}{n!\,(2r^\alpha)^n}
    + \c R^{(N+1)}_{3D}\bigg],\\
\label{e:l1dasymp}
&\c L_{1D}(r,\alpha) = \rec{\pi r}\bigg[\sum_{n=1}^N(-1)^{n+1}\m{sin}\Big(\frac{n\pi\alpha}2\Big)\frac{\Gamma(\alpha n{+}1)}{n!\,(2r^\alpha)^n}
    + \c R^{(N+1)}_{1D}\bigg],
\end{align}
where the residual $\c R$ terms have bounds that are similar to the terms in the sum:
\begin{align}
&\big|\c R_{3D}^{(N)}\big|\,{\le}\,\rec{\big[\m{sin}\big(\fract{\pi}{2\alpha}\big)\big]^{\alpha N+2}}\frac{\Gamma(\alpha N{+}2)}{N!\,(2r^\alpha)^N},\;\;\;\quad
\big|\c R_{1D}^{(N)}\big|\,{\le}\,\rec{\big[\m{sin}\big(\fract{\pi}{2\alpha}\big)\big]^{\alpha N+1}}\frac{\Gamma(\alpha N{+}1)}{N!\,(2r^\alpha)^N}.
\end{align}
\Eqs{e:l3dasymp}{e:l1dasymp} are thus asymptotic expansions: they can be used reliably for large $r$, where the first few terms are satisfactory (except for $\alpha\,{=}\,2$, when they do not say much). In particular, for $\alpha\,{\neq}\,2$, we see the power-law-like decrease, $\propto r^{-3-\alpha}$ for $\c L_{3D}$, $\propto r^{-1-\alpha}$ for $\c L_{1D}$. However, at no fixed $r$ does one get a convergent expansion.

There is thus an intermediate range of $r$ where we resorted to numerical integration. In the original \r{e:l13ddef} form this is numerically challenging due to fast oscillations, especially at larger $r$. This can be alleviated using complex analysis: in the \r{e:l13deiqr} form we can deform the integration contour into the upper half of the complex $q$ plane, so that the integrand is endowed with an exponential decay (from the $e^{iqr}$ factor). Practically, we can parametrize the path with $|q|\equiv t$ as
\begin{align}
q(t)=te^{i\varphi(t)} \Follows dq = dt\cdot e^{i\varphi(t)}\big[1 + it\varphi'(t)\big],
\end{align}
where we suppressed the $t$ dependence in the $\varphi(t)$ function. We thus arrive at
\begin{align}
\label{e:l3dphi}
&\c L_{3D}(r,\alpha) = \rec{2\pi^2r}\Im\sint{-2pt}{-12pt}0\infty\m dt\,te^{2i\varphi}\big[1\,{+}\,it\varphi'\big]e^{
irt\cos\varphi-rt\sin\varphi-\frac{t^\alpha}2\cos(\alpha\varphi)-i\frac{t^\alpha}2\sin(\alpha\varphi)},\\
\label{e:l1dphi}
&\c L_{1D}(r,\alpha) = \rec\pi\Re\sint{-2pt}{-12pt}0\infty\m dt\,e^{i\varphi}\big[1\,{+}\,it\varphi'\big]e^{
irt\cos\varphi-rt\sin\varphi-\frac{t^\alpha}2\cos(\alpha\varphi)-i\frac{t^\alpha}2\sin(\alpha\varphi)}.
\end{align}
We can choose the $\varphi(t)$ function optimally. We investigated two possibilities.
\vspace{1mm}\\
$A)$ One is requiring the modulus of the exponential factor to be as small as possible, for any $t$, to ensure an exponential suppression as fast as possible. We thus require
\begin{align}
\label{eq:kit}
e^{-rt\sin\varphi-\rec2t^\alpha\cos(\alpha\varphi)}\quad\tn{to be minimal}\Equivalent \cos\varphi=\alpha \fract{t^{\alpha-1}}{2r}\sin(\alpha\varphi) .
\end{align}
A satisfactory approximate solution can be found by expanding both sides in $\varphi$ up to second order around $\varphi{=}0$. We thus choose $\varphi(t)$ here accordingly as
\begin{align}
\varphi(t) = \sqrt{2\,{+}\,\alpha^4\fract{t^{2\alpha-2}}{4r^2}}\,{-}\,\alpha^2\fract{t^{\alpha-1}}{2r}.
\end{align}
Utilizing the power series of the sine and cosine functions   
we can estimate the decay of the modulus of the integrand in \Eqs{e:l3dphi}{e:l1dphi} with the chosen $\varphi(t)$ as
\begin{align*}
e^{-rt\sin\varphi(t)-\fract{t^\alpha}2\cos(\alpha\varphi(t))} \leq \m{exp}\bigg(
{-}rt\Big[\rect3\big(\alpha^4\fract{t^{2\alpha-2}}{4r^2}{+}2\big)^{3/2}{-}(\alpha^2{-}1)\fract{t^{\alpha-1}}{2r}{-}\rect3\alpha^6\fract{t^{3\alpha-3}}{8r^3} \Big]\bigg),
\end{align*}
with which we can determine the maximal $t$ value required for the integrand to vanish to a prescribed small value, e.g. $10^{-22}$ chosen in our calculations. In this way we obtained an integral that is numerically much more reliable for intermediate-large $r$ values than the original definitions in Eq.~\r{e:l13ddef}.
\vspace{2mm}\\
$B)$
Another possibility for choosing $\varphi(t)$ in \Eqs{e:l3dphi}{e:l1dphi} is to require the second exponential there to be a real positive function. (This condition, although one could think of more accurately demanding ones, is easily solvable, and we shall see below that a satisfactory formula is obtained.) Our requirement here is thus
\begin{align}
rt\cos\varphi - \rect2t^\alpha\sin(\alpha\varphi) = 0,
\end{align}
an equation that can be solved numerically for $\varphi$ on the fly, at any $t$ values. We denote this solution by $\varphi_t$; the derivative $\varphi'(t)$ can also be expressed using the rule of derivatives of implicit functions. After some simplification we get
\begin{align}
\label{e:l3dt}
&\c L_{3D}(r,\alpha) = \rec{2\pi^2r}\sint{-2pt}{-12pt}0\infty\m dt\,\frac{t\cos\varphi_t\big[\sin(\alpha\varphi_t)\,{+}\,\alpha\,\m{sin}\big((2{-}\alpha)\varphi_t\big)\big]}
{\sin(\varphi_t)\sin(\alpha\varphi_t)\,{+}\,\alpha\cos(\varphi_t)\cos(\alpha\varphi_t)}e^{{-}rt\frac{\cos((\alpha{-}1)\varphi_t)}{\sin(\alpha\varphi_t)}},\\
\label{e:l1dt}
&\c L_{1D}(r,\alpha) = \rec\pi\sint{-2pt}{-12pt}0\infty\m dt\,\frac{\alpha\cos(\varphi_t)\,\m{cos}\big((\alpha{-}1)\varphi_t\big)}
{\sin(\varphi_t)\sin(\alpha\varphi_t)\,{+}\,\alpha\cos(\varphi_t)\cos(\alpha\varphi_t)}e^{{-}rt\frac{\cos((\alpha{-}1)\varphi_t)}{\sin(\alpha\varphi_t)}}.
\end{align}
An even simpler-looking form can be obtained by transforming the integration variable to $x\,{=}\,t\cos\varphi_t$. This way we arrive at the following expressions, with
$\varphi_x$ implicitly determined by the $(\cos\varphi_x)^\alpha = \frac{x^{\alpha-1}}{2r}\sin(\alpha\varphi_x)$ condition: 
\begin{align}
&\c L_{3D}(r,\alpha) = \rec{2\pi^2r}\sint{-2pt}{-12pt}0\infty\m dx\,\frac{\sin(\alpha\varphi_x){+}\alpha\sin((2{-}\alpha)\varphi_x)}
{\alpha\cos\varphi_x\cos((\alpha{-}1)\varphi_x)}\cdot x\,\m{exp}\big({-}rx\fract{\cos((\alpha{-}1)\varphi_x)}{\cos\varphi_x\sin(\alpha\varphi_x)}\big),\\
&\c L_{1D}(r,\alpha) = \rec\pi\sint{-2pt}{-12pt}0\infty\m dx\,\m{exp}\big({-}rx\fract{\cos((\alpha{-}1)\varphi_x)}{\cos\varphi_x\sin(\alpha\varphi_x)}\big).
\end{align}
Numerically, the main advantage of these is that the integrand is a real positive monotonically decreasing function. In the study that follows, we used an assortment of the calculational methods expounded above, mainly for calculating $\c L_{1D}$ (the 1 dimensional projection); we gave the (similar) formulae for $\c L_{3D}$ for future reference.

\section{Source functions from the EPOS event generator}\label{s:application}

\subsection{Analysis method}

For our analysis we studied $\sqrt{s_{NN}}=200$ GeV Au+Au collisions generated by the EPOS program package. The EPOS model~\cite{Werner:2010aa} (Energy conserving quantum mechanical multiple scattering approach, based on Partons (parton ladders), Off-shell remnants, and Saturation of parton ladders) is a relatively freshly developed event generator used for modeling heavy-ion collisions. It divides the time evolution into different stages. The initial state is traced back to the parton model of the strongly interacting particles, then the interaction of the partons (quarks and gluons) is described based on the Lund String Model. After that comes a stage governed by viscous hydrodynamic expansion from the initial condition that is the outcome of the string model phase. The hadronization of the sQGP is modeled with well established fragmentation functions. After this, there comes an interacting hadronic gas state (with inelastic scatterings) described by the UrQMD model up until kinetic freeze-out. The EPOS event generator thus takes almost all of the important theoretical components of the description of heavy-ion collisions into account.

We were interested in the event-by-event geometry of the particle emitting source. For this analysis we considered only the most (0\%-5\%) central events. We investigated same charged final state pion pairs, in 3 classes of pair transverse momentum $K_T\,{\equiv}\,(K_x^2{+}K_y^2)^{1/2}$, with the total kinematic reach spanning $0.2\,\m{GeV}/c$ through $1.5\,\m{GeV}/c$ in terms of individual particle transverse momentum. We also made a cut on the individual particle pseudorapidity $\eta\,{\equiv}\,\rec2\ln\frac{p{+}p_z}{p{-}p_z}$ requiring $|\eta|\,{\leq}\,1$, resembling actual experimental conditions. For the study of the three-dimensional pair source distributions $D(\v r)$ we made histograms of the components of the pair relative coordinate $\v r$. We utilized the LCMS (longitudinally co-moving) coordinate system, in which the components of $\v r$ are expressed from the lab components, $\v r_{\m{lab}}$, as well as the $t_{\m{lab}}$ time difference of final particle creation as
\vspace{-5mm}\\
\begin{align}
 x=x_\m{lab}, \qquad  y=y_\m{lab}, \qquad z=\gamma(z_\m{lab}-\beta\cdot t_\m{lab})
\label{eq:LCMS}
\end{align}
\vspace{-5mm}\\
with $\gamma\,{=}\,(1{-}\beta^2)^{-1/2}$ and $\beta\,{\equiv}\,K_z/E$ with $E\,{\equiv}\,E_1{+}E_2$ being the total energy of the pair. In our directional study we chose the Bertsch-Pratt coordinates (directions) ,,long'', ,,out'' and ,,side'' (in the $z$ direction of the beam, in the direction of $K_T$, and perpendicular to these, respectively). Because the individual three-dimensional event-by-event histograms of $D(\v r)$ were so scarcely populated that inhibited meaningful fits, we took 1 dimensional projections of them and fitted these with the 1 dimensional L\'evy distribution (denoted by $\c L^{(1D)}(x_i,R_i,\alpha)$ in the previous section.
\vspace{-5mm}
\begin{figure}[H]
\begin{center}
\includegraphics[width=0.32\textwidth,trim={1.2mm 1.0mm 18.3mm 11mm},clip]{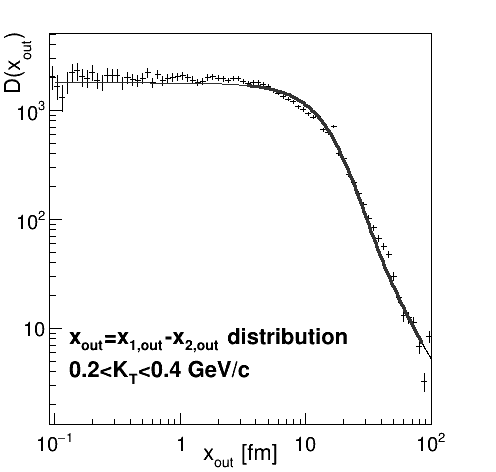}\hspace{0.01\linewidth}%
\includegraphics[width=0.32\textwidth,trim={1.2mm 1.0mm 18.3mm 11mm},clip]{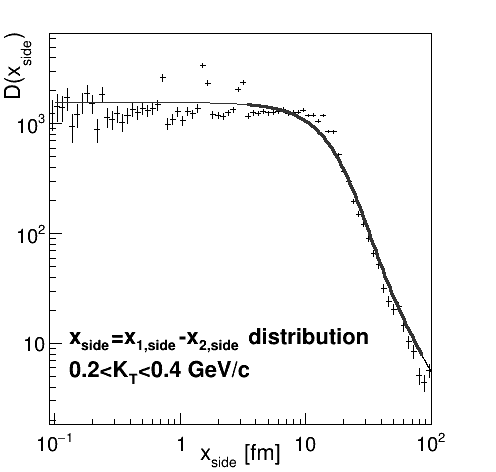}\hspace{0.01\linewidth}%
\includegraphics[width=0.32\textwidth,trim={1.0mm 1.0mm 18.3mm 11mm},clip]{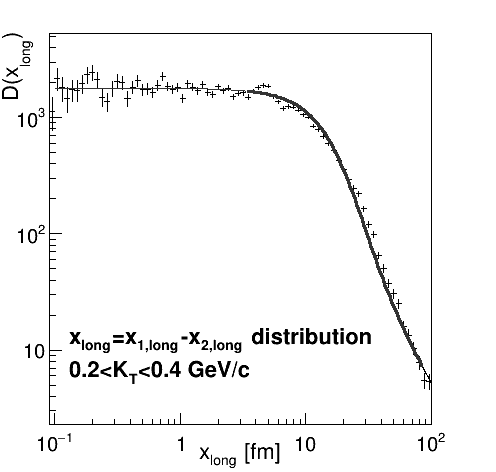}
\caption{
Example simultaneous fit for projections of the pair source function $D(\v r)$ in a given event and given $K_T$ region (plotted with doubly logarithmic axes to highlight the power-law-like asymptotics). The region included in the fit is denoted by thick lines.
}\label{fig:3_rszlsq}
\end{center}
\end{figure}\vspace{-10mm}

\subsection{Results}

An example of a fit of directional projections of L\'evy distributions to the three projections of the $D(\v r)$ histogram measured in one generated event can be seen on Fig.~\ref{fig:3_rszlsq}. The fitting for the three projections was done simultaneously with common L\'evy exponent (and common normalization) but different L\'evy scales.

\begin{figure}[H]
\includegraphics[width=0.45\textwidth,trim={2.0mm 1.0mm 18mm 11mm},clip]{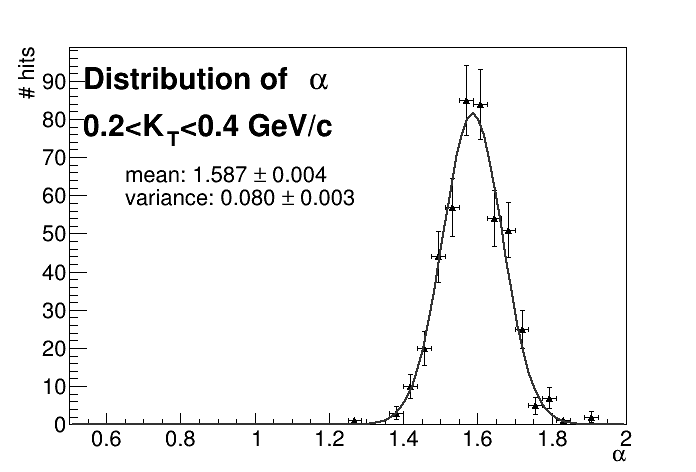}\hspace{0.02\linewidth}%
\includegraphics[width=0.45\textwidth,trim={2.0mm 1.0mm 18mm 11mm},clip]{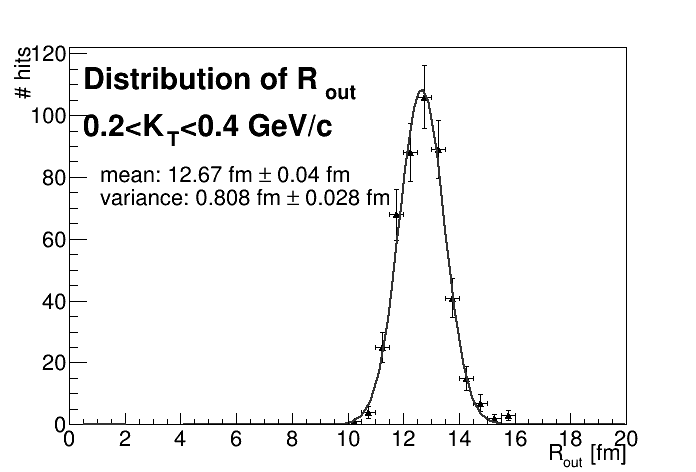}\\
\includegraphics[width=0.45\textwidth,trim={2.0mm 1.0mm 18mm 11mm},clip]{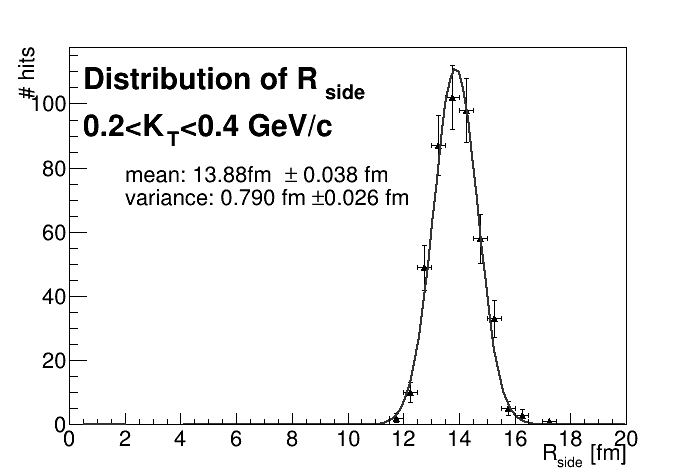}\hspace{0.02\linewidth}
\includegraphics[width=0.45\textwidth,trim={2.0mm 1.0mm 18mm 11mm},clip]{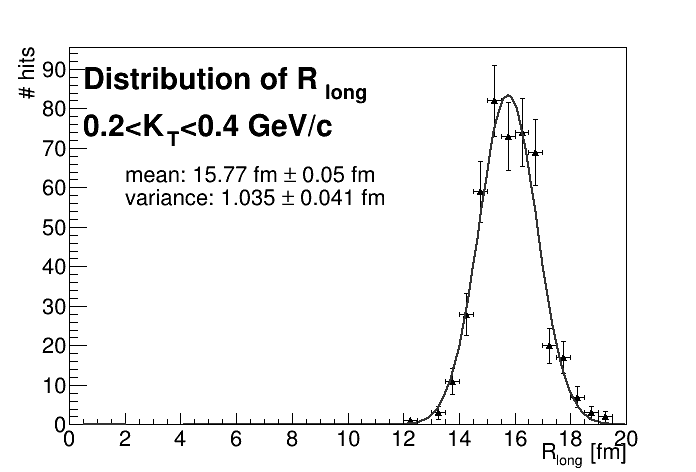}
\caption{
Distribution of the fitted parameters: L\'evy exponent $\alpha$ and the three directional L\'evy scales $R_\m{out}$, $R_\m{side}$ and $R_\m{long}$ along with Gaussians fitted to them, for events with centrality $0-5\%$ and pair transverse momentum range $0.4\,\m{GeV}/c\,{\leq}\,K_T\,{\leq}\,0.6\,\m{GeV}/c$.
}\label{fig:distr}
\end{figure}
\vspace{-6mm}
\begin{figure}[H]
\includegraphics[width=0.49\textwidth]{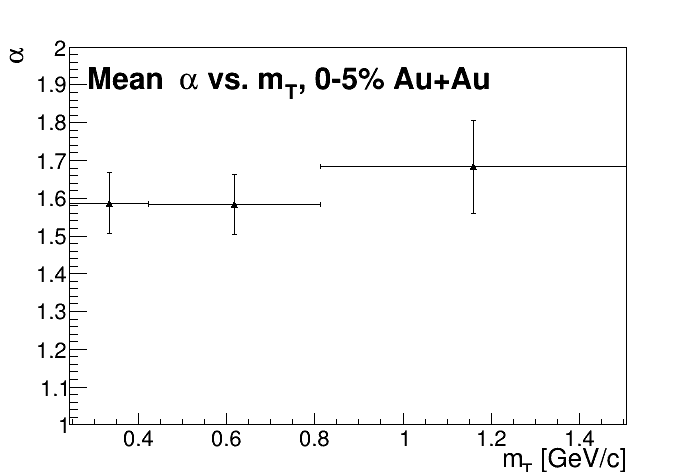}%
\includegraphics[width=0.49\textwidth]{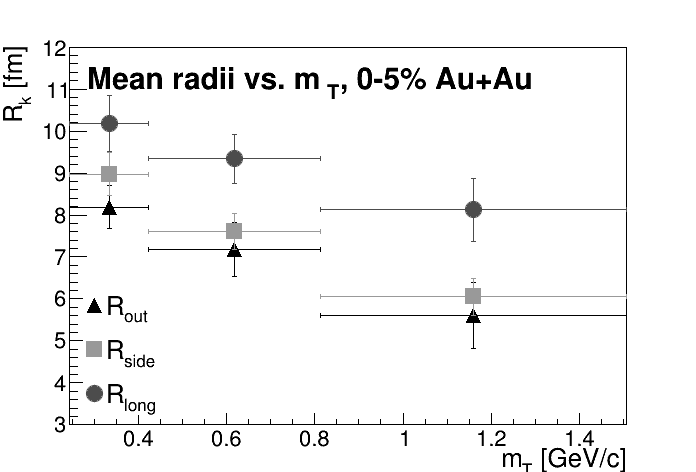}
\vspace{-1mm}
\caption{
Left: Event-by-event mean of the L\'evy exponent $\alpha$ vs. pair transverse momentum $K_T$. Values are around $1.6-1.7$, thus the Gaussian case (which would mean $\alpha\,{=}\,2$) is ruled out in this study. Right: event-by-event mean of the L\'evy scale values $R_\m{out}$, $R_\m{side}$ and $R_\m{long}$ vs. pair transverse momentum. There is a systematic difference hinting at small deviation from spherical symmetry, however, the general behaviors and magnitudes of the $R_k$ parameters match each other.}\label{fig:params}
\end{figure}

After doing these fits we studied the event-by-event distribution of the fit parameters $R_\m{out}$, $R_\m{side}$, $R_\m{long}$ and $\alpha$ in the three $K_T$ classes. An example distribution in a given transverse momentum region can be seen on Fig.~\ref{fig:distr}. These distributions in turn were fitted by Gaussians; the mean and variance of the result for the different parameters is then summarized in Table~\ref{tab:1}, and are visualized on Fig.~\ref{fig:params}.
\vspace{-5mm}\\
\begin{table}[H]
\centering
\tbl{}
{\begin{tabular}{|c|c|c|c|}
\hline
$K_T$   &   $0.2\,{\leq}\,K_T\,[\m{GeV}/c]\,{\leq}\,0.4$   &   $0.4\,{\leq}\,K_T\,[\m{GeV}/c]\,{\leq}\,0.6$   &   $0.6\,{\leq}\,K_T\,[\m{GeV}/c]\,{\leq}\,0.8$ \\ \hline
$\langle\alpha\rangle$           & 1.615  & 1.607 & 1.675 \\ \hline
$\sigma_{\alpha}$                & 0.085  & 0.115 & 0.224 \\ \hline
$\langle R_\m{out}\rangle$ [fm]  & 8.305  & 7.237 & 5.302 \\ \hline
$\sigma_{R_\m{out}}$ [fm]        & 0.515  & 0.713 & 1.534 \\ \hline
$\langle R_\m{side}\rangle$ [fm] & 8.902  & 7.486 & 5.952 \\ \hline
$\sigma_{R_\m{side}}$ [fm]       & 0.572  & 0.349 & 0.366 \\ \hline
$\langle R_\m{long}\rangle$ [fm] & 10.519 & 9.420 & 8.128 \\ \hline
$\sigma_{R_\m{long}}$ [fm]       & 0.931  & 0.638 & 0.863 \\ \hline
\end{tabular}
\caption{
Mean values and variances of the event-by-event fit results of the fit parameters for events with centrality 0-5\% and for different $K_T$ classes.\\\vspace{-2mm}
}\label{tab:1}}
\end{table}
\mbox{}\vspace{-6mm}\\
On Fig. \ref{fig:params} one can see that the L\'evy exponent $\alpha$ turned out to be around $1.6-1.7$ (roughly independent of transverse momentum $K_T$), significantly different from the Gaussian case of $\alpha\,{=}\,2$. This means that even without event and directional averaging, the appearance of L\'evy distributions is established in realistic simulations.
\section{Summary}\label{s:summary}

We presented a study of the objective of Bose-Einstein correlation measurements, the pairwise source distribution of particle emission in heavy-ion collisions using the EPOS event generator, posing the question if the experimentally observed appearance of L\'evy distributions instead of Gaussians is caused by event and/or directional averaging. We investigated the pair distribution determined by the event generator on an event-by-event basis utilizing three dimensional L\'evy distributions. As a digression, we discussed various methods of calculating such distribution functions reliably and effectively, a key ingredient that enabled our present work.

We studied the pair coordinate distributions of like sign final state pion pairs in 3 different transverse momentum classes, from 0-5\% most central $\sqrt{s_{NN}}=200$ GeV Au+Au collisions generated with EPOS event generator. Simultaneous event-by-event fits to the three projections of these distributions allowing different L\'evy scales (radius parameters) in the different directions revealed that even without directional and event-by-event averaging, an average L\'evy exponent significantly different from 2.0, namely $\alpha=1.6-1.7$ is best suited to describe individual event-by-event source functions. It is noteworthy that experimental results from Bose-Einstein correlation measurements also favor a significant $\alpha\,{<}\,2$ trend (i.e. non-Gaussianity), albeit different general value of $\alpha$~\cite{PHENIX:2017ino}. This might hint at interesting questions pertaining to different physical processes that play a role in the creation of such long-range components in the pion source function. The values of the three L\'evy scales (corresponding to the three different directions) are different from each other for every transverse momentum range: biggest in the beam direction, the other two being close to equal, however $R_\m{side}$ seems to be somewhat bigger than $R_\m{out}$. This is also interesting, as many hydrodynamical models predict the opposite ordering. It is also interesting on its own that the apparent source size decreases with transverse momentum in a very similar way to what is observed in hydrodynamical models, although the latter almost always assume (and predict) Gaussian sources.

A possible future generalization of our work is azimuthally sensitive analysis, to see the ultimate directionally resolved source function. We have, however, confidence now that the non-Gaussian source functions in high energy heavy-ion collisions is, instead of being only a measurement artifact, a phenomenon that carries deep physical meaning, to be explored by future measurements and model studies.

We thank D. Kincses and M. Csan\'ad for insightful discussions. This work was supported by the Hungarian NKFIH grants TKP2021-NKTA-64 and K-138136.


\begin{thebibliography}{99} 

\bibitem{PHENIX:2004vcz}
PHENIX Collab. (K.~Adcox \textit{et al}.) {\textit{ Nucl. Phys. A}\/} {\bf{757}}, 184-283 (2005).

\bibitem{STAR:2005gfr}
STAR Collab. (J.~Adams \textit{et al.}) \textit{Nucl. Phys. A} \textbf{757}, 102-183 (2005).

\bibitem{PHOBOS:2004zne}
PHOBOS Collab. (B.~B.~Back \textit{et al.}) \textit{Nucl. Phys. A} \textbf{757}, 28-101 (2005).

\bibitem{BRAHMS:2004adc}
BRAHMS Collab. (I.~Arsene \textit{et al.}) \textit{Nucl. Phys. A} \textbf{757}, 1-27 (2005).

\bibitem{PHENIX:2004yan}
PHENIX Collab. (S.~S.~Adler \textit{et al.}) \textit{Phys. Rev. Lett.} \textbf{93}, 152302 (2004).

\bibitem{PHENIX:2006nml}
PHENIX Collab. (S.~S.~Adler \textit{et al.}) \textit{Phys. Rev. Lett.} \textbf{98}, 132301 (2007).

\bibitem{PHENIX:2007grx}
PHENIX Collab. (S.~Afanasiev \textit{et al.}) \textit{Phys. Rev. Lett.} \textbf{100}, 232301 (2008).

\bibitem{PHENIX:2017ino}
PHENIX Collab. (A.~Adare \textit{et al.}) \textit{Phys. Rev. C} \textbf{97}, 064911 (2018) no.6

\bibitem{Kincses:2024sin}
D.~Kincses [STAR], \textit{Universe} \textbf{10}, 102 (2024) no.3

\bibitem{CMS:2023xyd}
CMS Collab. (A.~Tumasyan \textit{et al.}) \textit{Phys. Rev. C} \textbf{109}, 024914 (2024) no.2

\bibitem{NA61SHINE:2023qzr}
H.~Adhikary \textit{et al.} [NA61/SHINE],
Eur. Phys. J. C \textbf{83}, no.10, 919 (2023).

\bibitem{PHENIX:2024vjp}
PHENIX Collab. (N.~J.~Abdulameer \textit{et al.}) 
[arXiv:2407.08586 [nucl-ex]].

\bibitem{Csanad:2024hva}
M.~Csan\'ad and D.~Kincses, Universe \textbf{10}, no.2, 54 (2024).

\bibitem{Cimerman:2019hva}
J.~Cimerma\svejk n, C.~Plumberg and B.~Tom\'a\svejk sik, Phys. Part. Nucl. \textbf{51} (2020) no.3, 282-287.

\bibitem{Kincses:2022eqq}
D.~Kincses, M.~Stefaniak and M.~Csan\'ad, \textit{Entropy} \textbf{24}, 308 (2022) no.3

\bibitem{Korodi:2022ohn}
B.~K\'orodi, D.~Kincses and M.~Csan\'ad, Phys. Lett. B \textbf{847}, 138295 (2023).

%
\bibitem{Goldhaber:1960sf}
G.~Goldhaber, S.~Goldhaber, W.~Y.~Lee and A.~Pais, \textit{Phys. Rev.} \textbf{120}, 300-312 (1960).

\bibitem{Csorgo:1999sj}
T.~Cs\"org\H o, \textit{Acta Phys. Hung. A} \textbf{15}, 1-80 (2002).

\bibitem{Lisa:2005dd}
M.~A.~Lisa, S.~Pratt, R.~Soltz and U.~Wiedemann, \textit{Ann. Rev. Nucl. Part. Sci.} \textbf{55}, 357-402 (2005).

\bibitem{Yano:1978gk}
F.~B.~Yano and S.~E.~Koonin, Phys. Lett. B \textbf{78}, 556-559 (1978).

\bibitem{Csorgo:2003uv}
T.~Cs\"org\H o, S.~Hegyi and W.~A.~Zajc, \textit{Eur. Phys. J. C} \textbf{36}, 67-78 (2004).

\bibitem{Csorgo:2005it}
T.~Cs\"org\H o, S.~Hegyi, T.~Nov\'ak and W.~A.~Zajc, \textit{AIP Conf. Proc.} \textbf{828}, 525-532 (2006).

\bibitem{Csorgo:2004sr}
T.~Cs\"org\H o, S.~Hegyi, T.~Nov\'ak and W.~A.~Zajc, \textit{Acta Phys. Polon. B} \textbf{36}, 329 (2005).

\bibitem{Csanad:2007fr}
M.~Csan\'ad, T.~Cs\"org\H o, M.~Nagy, Braz. J. Phys. \textbf{37}, 1002 (2007).

\bibitem{Levy}
K.~Arias-Calluari, F.~Alonso-Marroquin and M.~S.~Harr\'e, \textit{Phys. Rev. E} \textbf{98}(1-1), 012103-012103 (2018).

\bibitem{Levy:old}
J.~P.~Nolan, \textit{Communications in Statistics. Stochastic Models}, \textbf{13}(4), 759–774 (1997).

\bibitem{Levy:num}
Y.~Liang and W.~Chen, \textit{Signal Processing} \textbf{93}, 242-251 (2013).

\bibitem{Werner:2010aa}
K.~Werner, I.~Karpenko, T.~Pierog, M.~Bleicher and K.~Mikhailov, \textit{Phys. Rev. C} \textbf{82}, 044904 (2010).

\end{thebibliography}
\end{document}